\documentclass[10pt, journal, letterpaper]{IEEEtran}
\usepackage{amsmath,amssymb,amsfonts}
\usepackage{algorithmic}
\usepackage{graphicx}
\usepackage{textcomp}
\usepackage{wrapfig}
\def\BibTeX{{\rm B\kern-.05em{\sc i\kern-.025em b}\kern-.08em
    T\kern-.1667em\lower.7ex\hbox{E}\kern-.125emX}}
\usepackage{float}
\usepackage{amsmath}
\usepackage{amssymb}
\usepackage{algorithmic}
\usepackage{graphicx}
\usepackage{textcomp}
\usepackage{xcolor}
\definecolor{subsectioncolor}{RGB}{0, 0, 255} % Defines subsectioncolor as blue
\usepackage{verbatim }
\usepackage{siunitx}

\usepackage{booktabs}
\usepackage{hyperref}
\usepackage{balance}
\usepackage{url}
\usepackage{tabularx}
\usepackage{array}
\usepackage{scalerel}
\usepackage{tikz}
\usetikzlibrary{svg.path}
\definecolor{orcidlogocol}{HTML}{A6CE39}
\tikzset{
  orcidlogo/.pic={
    \fill[orcidlogocol] svg{M256,128c0,70.7-57.3,128-128,128C57.3,256,0,198.7,0,128C0,57.3,57.3,0,128,0C198.7,0,256,57.3,256,128z};
    \fill[white] svg{M86.3,186.2H70.9V79.1h15.4v48.4V186.2z}
                 svg{M108.9,79.1h41.6c39.6,0,57,28.3,57,53.6c0,27.5-21.5,53.6-56.8,53.6h-41.8V79.1z M124.3,172.4h24.5c34.9,0,42.9-26.5,42.9-39.7c0-21.5-13.7-39.7-43.7-39.7h-23.7V172.4z}
                 svg{M88.7,56.8c0,5.5-4.5,10.1-10.1,10.1c-5.6,0-10.1-4.6-10.1-10.1c0-5.6,4.5-10.1,10.1-10.1C84.2,46.7,88.7,51.3,88.7,56.8z};
  }
}
\newcommand\orcidicon[1]{\href{https://orcid.org/#1}{\mbox{\scalerel*{
\begin{tikzpicture}[yscale=-1,transform shape]
\pic{orcidlogo};
\end{tikzpicture}
}{|}}}}

\usepackage{svg}

\begin{document}
\title{
Prediction of the Received Power of Low-Power Networks Using Inertial Sensors
}

\author{Waltenegus Dargie\orcidicon{0000-0002-7911-8081}, \IEEEmembership{Senior Member, IEEE}, Christian Poellabauer\orcidicon{0000-0002-0599-7941}, \IEEEmembership{Senior Member, IEEE}, and Abiy Tasissa\orcidicon{0000-0003-4033-7735}  
    \thanks{Manuscript submitted on 1 January 2025.}
    \thanks{This work has been partially funded by the German Research Foundation (DFG) under project agreements DA 1211/7-1 and NSF DMS 2208392}
    \thanks{W. Dargie is with the Faculty of Computer Science, Technische Universit{\"a}t Dresden, 01062 Dresden, Germany (e-mail: waltenegus.dargie@tu-dresden.de)}
    \thanks{C. Poellabauer is with the Knight Foundation School of Computing and Information Sciences at Florida International University, USA, (e-mail:  cpoellab@fiu.edu)}
    \thanks{A. Tasissa is with the Department of Mathematics, Tufts University, USA (Abiy.Tasissa@tufts.edu)} 
}
\maketitle

\begin{abstract}
Low-power and cost-effective IoT sensing nodes enable scalable monitoring of different environments. Some of these environments impose rough and extreme operating conditions, requiring  continuous adaptation and reconfiguration of physical and link layer parameters. In this paper, we closely investigate the stability of the wireless links established between nodes deployed on the surface of different water bodies and propose a model to predict the received power. Our model is based on Minimum Mean Square Estimation (MMSE) and relies on the statistics of received power and the motion the nodes experience during communication. One of the drawbacks of MMSE is its reliance on matrix inversion, which is at once computationally expensive and difficult to implement with resource constrained devices. We forgo this stage by estimating model parameters using the gradient-descent approach, which is much simpler to implement. The model achieves a prediction accuracy of 91\% even with a small number of iterations and a significant amount of water motion during communication.
\end{abstract}


\begin{IEEEkeywords}
Adaptive transmission power, Inertial Measurement Unit (IMU), wireless sensor networks, link quality estimation, water quality monitoring, Internet-of-Things 
\end{IEEEkeywords}


\section{Introduction}
\label{intro}

Cost-effective and low-power IoT nodes enable scalable and distributed sensing and have a wide range of applications \cite{lombardo2017wireless, zhao2024missing}. Some of these applications---water quality monitoring, pollution detection, flooding monitoring, and seaweed monitoring --- require the deployment of the nodes on the surface of rough waters, which affect the networks in a number of ways. Firstly, it constantly modifies their topology, thereby making the establishment of stable routes a significant challenge. Secondly, it makes the wireless links unstable, causing considerable packet loss. Thirdly, repeated attempts to reestablish links and retransmit packets can prematurely exhaust a device's energy resources \cite{gong2021self}. 

Dynamic transmission power adaptation is one of the mechanisms employed to deal with adverse external factors \cite{lin2016atpc, jurdak2007adaptive, dargie2011dynamic}. Often, it is jointly used with other mechanisms, such as dynamic channel selection \cite{watteyne2009reliability}. 
Most existing or proposed approaches on transmission power adaptation rely on received power statistics. The underlying assumption is that (1) given an appropriate path loss model, the relationship between the received power and the transmission power can be established and (2) the received power exhibits strong autocorrelation, which makes it predictable  \cite{govindan2011probability,alsamhi2021predictive}. This assumption holds for many real world situations (as is the case in cellular networks, for example). For the deployment environments we are concerned with, however, it is either partially true or may not be applicable at all. The reason is that the main cause of change in the received power (the motion of the water) is 3-dimensional whereas the effect is one-dimensional. The rougher the motion, the more complex is the relationship between the cause and the effect. A more plausible approach is to predict the received power from the motion of the water and to undertake compensatory measures (increase or decrease transmission power) to obtain good quality of service. The purpose of this paper is to propose an approach to achieve this goal. 

\begin{figure}
	\centering
		\includegraphics[width=0.450\textwidth]{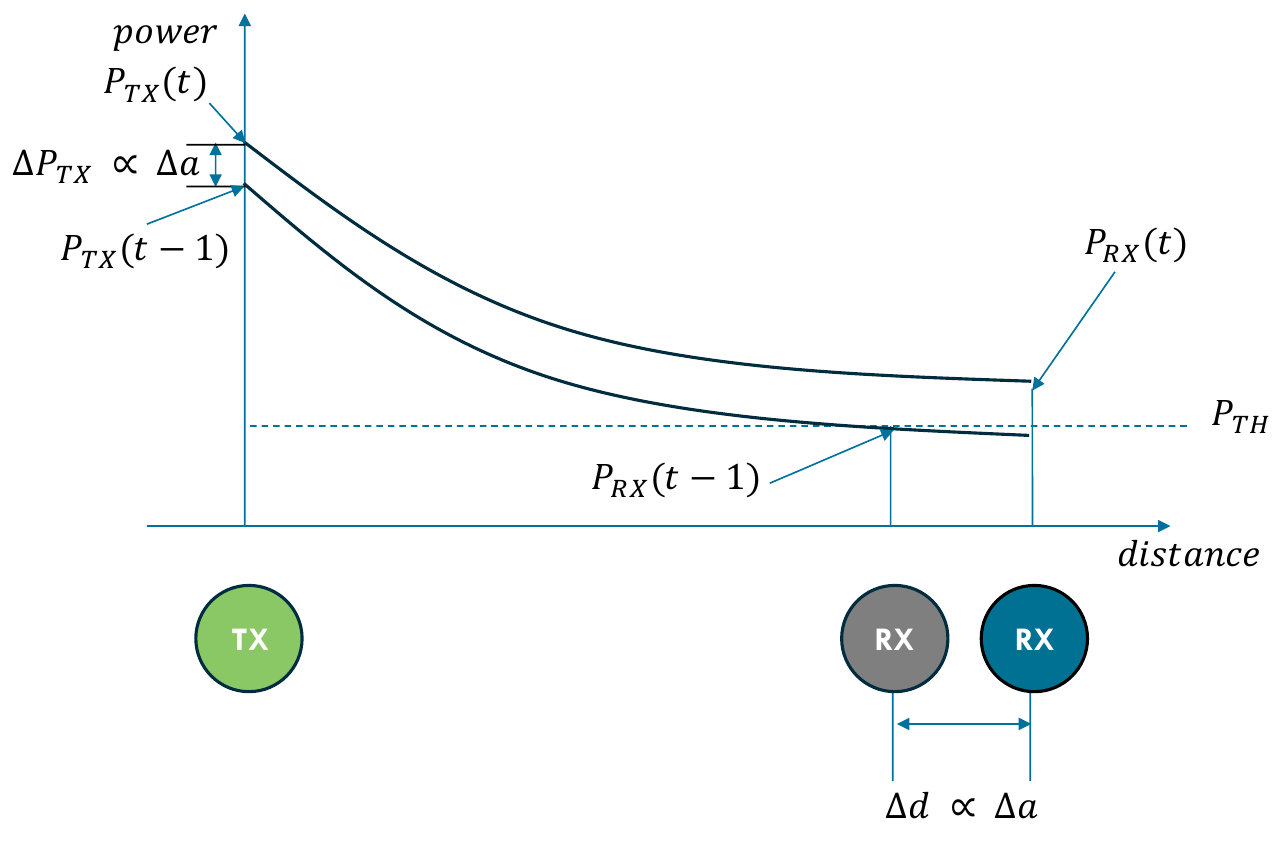}
	\caption{Illustration of dynamic power adaptation. A transmitter aims to transmit a packet with a power such that when received, the packet has a received power above a set threshold.}
	\label{fig:received_power_illustration}
\end{figure}

What we wish to achieve in this paper is illustrated in Figure~\ref{fig:received_power_illustration}: At a receiver, a threshold power is set for received packets to be successfully decoded. A transmitter predicts the received power based on a locally available, lightweight model and adjusts the power of outgoing packets, so that when they are received, their power is above the set threshold. As the first contribution of this paper, we express the received power in terms of the 3-dimensional acceleration the nodes experience and describe the relationship between these quantities (defined as random variables) using Minimum Mean Square Estimation (MMSE) \cite{papoulis2002probability}. However, MMSE involves matrix inversion, which is computationally expensive and difficult to implement in resource-constrained devices. Therefore, we forego matrix inversion in favor of gradient-descent, which is much less complex to implement. Though gradient-descent involves iterative estimation and matrix multiplication, we shall experimentally demonstrate that with very few iterations, the model achieves a remarkable accuracy. As the second contribution, with practical deployments on different water bodies using resilient prototypes and two different heterogeneous low-power radios, we demonstrate the reliability and accuracy of our model. More specifically, the model achieves on average a prediction accuracy exceeding 91\%, which is remarkable for resource-constrained devices and very rough deployment environments. 

The remaining part of the paper is organized as follows. In the following section, we describe related work. In Section \ref{sec:deployment}, we describe the deployment scenarios wherein we conducted extensive experiments. In Section \ref{sec:model}, we introduce our approach to model low-power and lossy wireless links. In Sections~\ref{sec:evaluation} and~\ref{sec:comp} we present test results, and compare them with the state-of-the-art. Finally, in Section \ref{sec:conclusion}, we provide concluding remarks and outline future work.

\section{Related Work}
\label{sec:related}

Recent research on wireless radio performance in extreme environments has explored various innovative technologies and techniques to enhance wireless communications \cite{Renzone9361700}. This includes the use of multi-radio, multichannel wireless radios to improve overall network performance~\cite{10.1109/mascot.2009.5366718}, cognitive radios to intelligently optimize spectrum utilization~\cite{10.1155/2018/4173810}, and machine learning techniques to dynamically optimize wireless network performance, as proposed in~\cite{giordano2022design, 10.1587/comex.2018xbl0061}. Environmental conditions such as weather, terrain, and human activities have been widely recognized for their impacts on radio performance such as fluctuations in transmission rates and quality of service in wireless sensor networks, e.g., the work in~\cite{10.5194/acp-9-2413-2009} investigated the impact of weather conditions and atmospheric phenomena on link quality, while other research~\cite{10.1145/1089444.1089466} studied the impacts of humidity and movement, all of which can affect radio signal quality. Wireless devices, including wireless sensor networks, deployed near water surfaces also experience a variety of quality fluctuations, e.g., the absorption and scattering of water and small particles dissolved in it have been found to significantly affect wireless optical signals~\cite{10.18280/mmep.080316}. In~\cite{10.1016/j.jart.2017.07.004} and~\cite{10.3390/s140916932}, the authors survey various applications of wireless sensor networks for the real-time monitoring of water quality, marine life, and environmental conditions in marine areas. While focusing primarily on underwater scenarios, the work in~\cite{10.1155/2017/7539751} discusses specifically the challenges of energy-efficiency, topology control, and transmission power control strategies. The conductive properties of the water surface can strengthen signal reflections and increase interference effects, as discussed in~\cite{9706046}. Similar to our own observations, this work also observed that recurrent natural phenomena, such as tides or waves, lead to water level variations and therefore changing interference patterns, affecting the signal propagation over water surfaces. The focus of the work in~\cite{9706046}, however, is on determining the optimal combination of radio height and distance that minimizes the path loss averaged during a whole tidal cycle. In contrast, in our work, we assume that devices are floating on the surface of the water, and we address link quality changes by predicting future received powers, which can be used to design an adaptive transmit power scheme. Finally, in~\cite{9894283}, the authors also attempt to predict path loss for a Long Range (LoRa) line-of-sight link deployed over an estuary with characteristic intertidal zones, studying both shore-to-shore and shore-to-vessel communications. While this work does not employ a transmit power adaptation scheme to address path loss variations, the proposed technique could certainly be used to help decide on future transmit powers. In contrast to the work in~\cite{9894283}, the focus of our work is on wireless sensor networks, where devices are deployed directly on the surface of a body of water (instead of shore- or vessel-based radios).

\section{Deployment}
\label{sec:deployment}

\begin{figure*}
	\centering
	\includegraphics[width=0.9\textwidth]{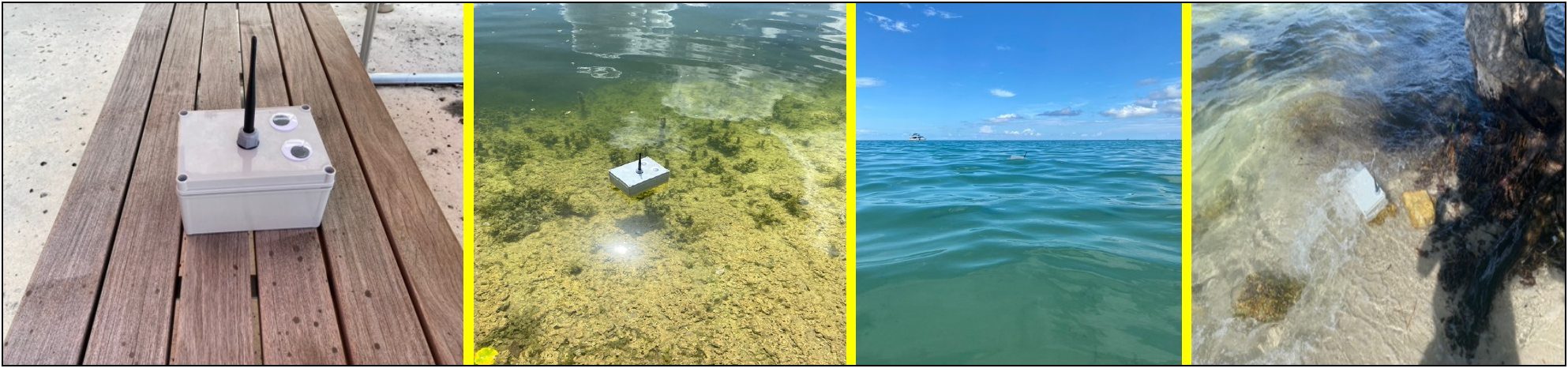}
	\caption{The deployment of waterproof buoys on the surface of different water bodies. From left to right: A prototype low-power IoT node; deployments on: a lake on FIU's Modesto A. Maidique Campus, Miami South Beach, and Crandon Beach.}
	\label{fig:deployments}
\end{figure*}

In order to study link quality fluctuation, we deployed wireless sensor networks on four different water bodies in Miami, Florida, between the first of June and the end of August 2023. The first deployment was on one of the small lakes on the main campus of Florida International University (FIU). The second was on North Biscayne Bay; and the third and the fourth were on South Beach, Miami, and Crandon Beach, Miami, respectively. Some of the deployments took place at the time when the State of Florida was significantly affected by Hurricane Idalia, a  Category 4 hurricane. The first two water bodies were relatively calm, and the main causes of link quality fluctuation were heavy rain and excessive heat. The last two, on the other hand, besides experiencing extreme weather conditions, moved significantly, causing frequent disconnections of the wireless links. We placed the sensor nodes in waterproof boxes and used waterproof marine antennas to transmit and receive electromagnetic signals in the sub-Giga and 2.4 GHz frequency bands (the sensor nodes integrate two different radio chips, CC1200 and CC2538). In order to establish a relation between the change in RSSI (signifying link quality fluctuation) and the movement of water, we also embedded 3D accelerometers and gyroscopes in the waterproof buoys. Figure~\ref{fig:deployments} shows the deployments of the buoys in the three different environments.

\section{Model}
\label{sec:model}

The two radios integrated into our sensor nodes -- the CC1200 \cite{instruments2013cc1200} and the CC2538 System-on-Chip \cite{instruments2013cc2538} -- support dynamic transmission power adaptation and define thresholds for successfully receiving packets. The CC1200 has a sensitivity of -123 dBm and its transmission power can be adjusted up to a maximum level of 16 dBm. This radio operates in sub-Giga Hertz frequency bands. The CC2538 radio chip operates in the 2.4 GHz and has a sensitivity of -97 dBm. Its transmission power can be adjusted up to a maximum level of 7 dBm. In addition, the Contiki operating system  \cite{oikonomou2022contiki} -- which we used to set up the sensing network -- enables the definition of soft thresholds to improve the reliability of packet reception. 

Once a receiving node defines the threshold power of incoming packets, the purpose of an adaptive transmission power model is to estimate the transmission power, so that packets reaching the receiving node are received with a minimum power that equals or exceeds the threshold power. This task requires a feedback system and can be accomplished in two ways. The decision leading to dynamic power adaptation can be made either at the transmitter or at the receiver. In the first case, the transmitter evaluates the power with which ACK packets are received and estimates how far the receiver is from it and adjusts the transmission power of the next packet accordingly. In the second, the decision is made at the receiver; by evaluating the power of received packets, it adjusts its own power to transmit ACK packets. The power it uses implicitly reminds the transmitter to adjust its transmission power for the next packet. In either case, a lightweight model is required to predict the received power.

The relationship between the transmitted power, received power, and the distance of separation for a line-of-sight (LOS) wireless link is classically expressed as follows: 
\begin{equation}
\label{eq:model_1}
P_{rx} \propto G_{tx} G_{rx} A_{tx} A_{rx} \frac{P_{tx}}{d^2},
\end{equation}
where $G$ and $A$ refer to the gains and the cross-sectional areas of the transmitter and the receiver antennas. The terms can be cleanly separated if the received power is expressed in decibel:
\begin{equation}
\label{eq:model_2}
10 \; log \left (P_{rx} \right ) =  K + 10 \; log \left ( P_{tx} \right ) - 20 \; log \left (d \right ),
\end{equation}
where $K = 10 \; \log \left (G_{tx} G_{rx} A_{tx} A_{rx} \right) $ is a constant once the antennas are selected and can be experimentally determined by fixing $P_{tx}$ and $d$ and observing the received power. For example, for the CC2538 radio, at $10m$ distance and 0 dBm transmission power, the received power is on average -50 dBm (consequently, $K = 10^{-3}$). Once $K$ is established, the received power can be expressed in terms of the transmission power and the distance. The advantage of this model is that its computational cost is modest. For our case, both radios provide the power of received packets in dBm and their transmission powers are configured in dBm. Nevertheless, the usefulness of this model is limited on account of the constant and complex motion the communicating nodes undergo. In addition, the LOS assumption may not hold, or holds only marginally, since  water scatters, refracts, and reflects the electromagnetic signal \cite{van2013propagation}.

Can the received power be predicted from its immediate past history? This can be answered in the affirmative if the received power exhibits a strong autocorrelation. Once we have sufficient statistics of the received power, it is possible to test this. The advantage of this approach is that the statistics can be established and updated from received packets alone. One of its limitations is that due to the complexity of the motion  the nodes undergo, the correlation between neighbor samples could be too weak. 

Alternatively, a direct correlation between the underlying motion and the link quality fluctuation can be established. This can be achieved in two ways. One of them is to estimate the relative distance of the receiver (i.e., relative to its past position) as a function of the perceived change in the acceleration, $\Delta d =  f( \Delta \textbf{a})$. Once $\Delta d$ is estimated, it can be summed to the previously known displacement of the node in order to determine the overall displacement, $d$. Then $d$ can be inserted in Equation \ref{eq:model_1} to estimate the transmission power which satisfies the threshold condition. This approach, too, is error-prone due to the difficulty of estimating displacement from the 3D acceleration. Alternatively, the change in the received power can be directly correlated with the change in the perceived acceleration, so that $\Delta P_{rx}$ can be expressed in terms of $ f(\Delta \textbf{a})$.  

\subsection{Linear Relationship}
Given the relationship between two random variables $\textbf{x}$ and $\textbf{y}$, it is possible to establish the statistics of one of the random variables in terms of the statistics of the other random variable. For example, if $\textbf{y} =  a$\textbf{x}$ + b$, where $a$ and $b$ are constants and $a > 0$, then the distribution of $\textbf{y}$ can be established in terms of the distribution of $\textbf{x}$:
\begin{align}
    \label{eq:rel_1}
    F_y(y) & = P\left \{ \mathbf{y} \leq y \right \} = P\left \{ \left (a\mathbf{x} + b  \right ) \leq y \right \} \\ \nonumber
           & = P\left \{ \mathbf{x} \leq \frac{y - b}{a}  \right\}
           =  F_x \left (  \frac{y - b}{a}  \right ).
\end{align}

\begin{figure}
	\centering
	\includegraphics[width=0.45\textwidth]{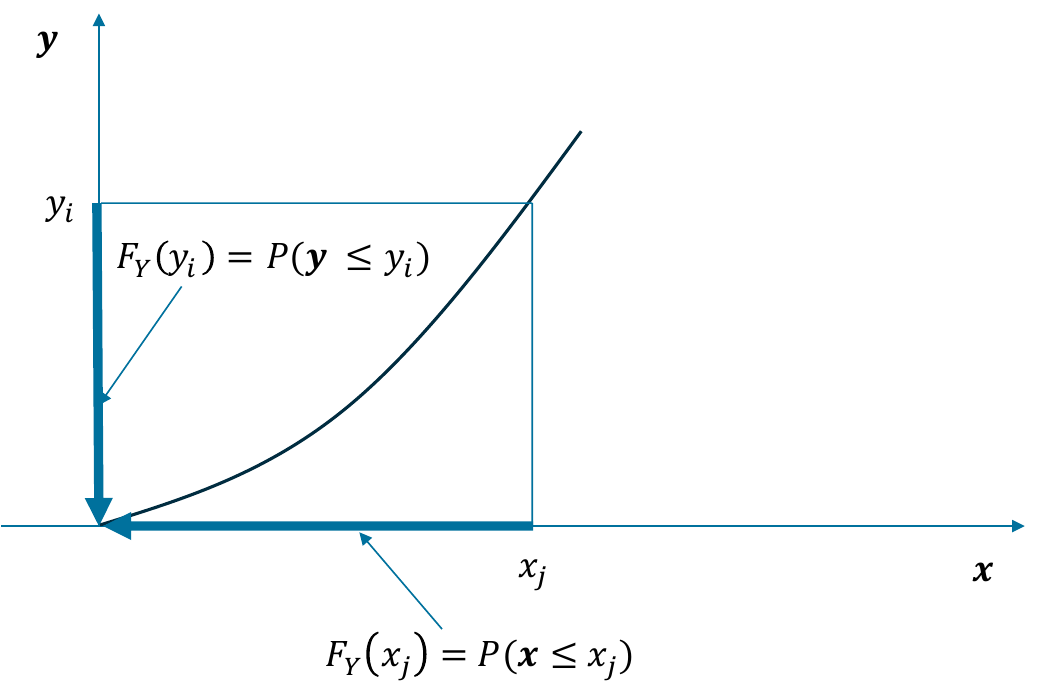}
	\caption{Relationship between two random variables.}
	\label{fig:random_variables}
\end{figure}

Similarly, if two random variables have a one-to-one relationship, their relationship can be estimated from their distribution functions, assuming that these are available. Consider Figure ~\ref{fig:random_variables}; given that the random variable $\mathbf{y}$ is the dependent variable (and $\mathbf{x}$, the independent variable), then it is true that for every $y_i$, there is a corresponding $x_j$. Furthermore,
\begin{equation}
    \label{eq:rel_2}
     P\{\mathbf{y} \leq y_i \} = P\{\mathbf{x} \leq x_j \}.
\end{equation}
In other words,
\begin{equation}
    \label{eq:rel_3}
    F_y(y_i) = F_x(x_j).
\end{equation}
From which we can derive (ref. to \cite{dargie2014stochastic}):
\begin{equation}
    \label{eq:rel_4}
    y_i = F^{-1}_y \left (F_x(x_j) \right ).
\end{equation}

An interesting aspect of Equation~\ref{eq:rel_3} is that if the two distributions have similar characteristics, we can suspect the existence of a linear relationship between $\mathbf{x}$ and $\mathbf{y}$. We shall demonstrate this for two distribution functions, uniform and exponential. Suppose both $\mathbf{x}$ and $\mathbf{y}$ are uniformly distributed, the former between $[0, a]$, and the latter between $[0,b]$. Hence, $F_y(y) = y/a$ and $F_x(x) = x/a$, from which we conclude (applying Equation~\ref{eq:rel_4}): $y = (b/a) x$. Now suppose that the two random variables are exponentially distributed, with $F_y(y) = ( 1 -  e^{-\lambda_y y})$ and   $F_x(x) = (1 - e^{-\lambda_x x}), \lambda_x, \lambda_y > 0$.  Applying Equation~\ref{eq:rel_4} to these distribution functions leads to:
\[
\mathbf{y} = \left [ \frac{\lambda_x}{\lambda_y} \right ]  \mathbf{x}.
\]
 Since $\lambda_x$ and  $\lambda_y$ are constants, so is their ratio, so that: 
\[
\mathbf{y} = A \mathbf{x}.
\]
where $A = \lambda_x/\lambda_y$. The assertion of a linear relationship is valid for any type of distributions.

\begin{figure}
	\centering
	\includegraphics[width=0.48\textwidth]{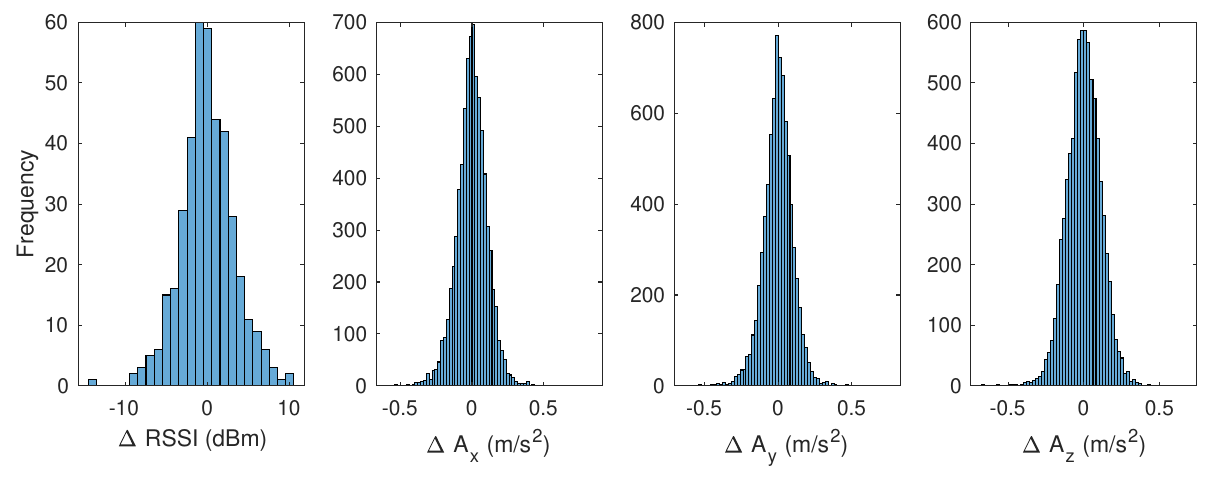}
	\caption{Histograms depicting the change in the RSSI of received packets and the change in the linear acceleration of the transmitting node. The nodes were using the CC1200 radio and the channel had a center frequency of 869.3 MHz and a bandwidth of 2 MHz.}
	\label{fig:histogram_cc1200}
\end{figure}

\begin{figure}
	\centering
	\includegraphics[width=0.48\textwidth]{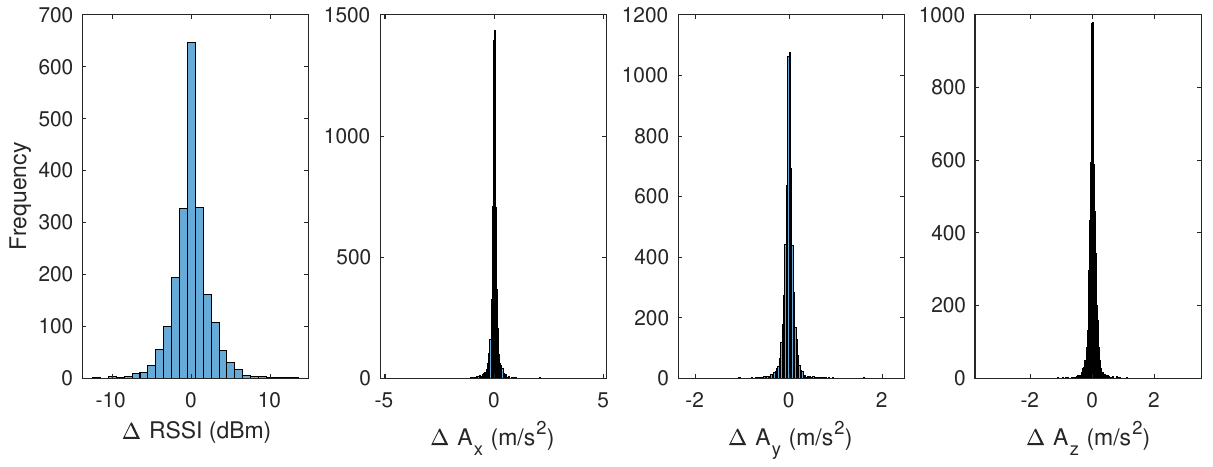}
	\caption{Histograms depicting the change in the RSSI of received packets and the change in the linear acceleration of the transmitting node. The nodes were using the CC2538 radio (2.4 GHz).}
	\label{fig:histogram_cc2538}
\end{figure}

Figures \ref{fig:histogram_cc1200} and~\ref{fig:histogram_cc2538} display the histograms of the changes in the RSSI and linear acceleration for the CC2538 and CC1200 radios and the deployments we carried out at Crandon Beach, Miami. The statistics are established based on 5 independent experiments carried out on different days. Each experiment lasted approximately 30 minutes. The pattern is similar for the deployments carried out at South Beach and North Biscayne Bay. This suggests that the change in the received power can be regarded as linearly correlated with the changes in the linear accelerations. If this observation is valid, then it is possible to employ linear models to establish a relationship between $\Delta P_{tx}$ and $f\Delta \mathbf{a}$.

\subsection{Minimum Mean Square Estimation}
\label{sec:mmse}

\begin{figure}
	\centering
	\includegraphics[width=0.48\textwidth]{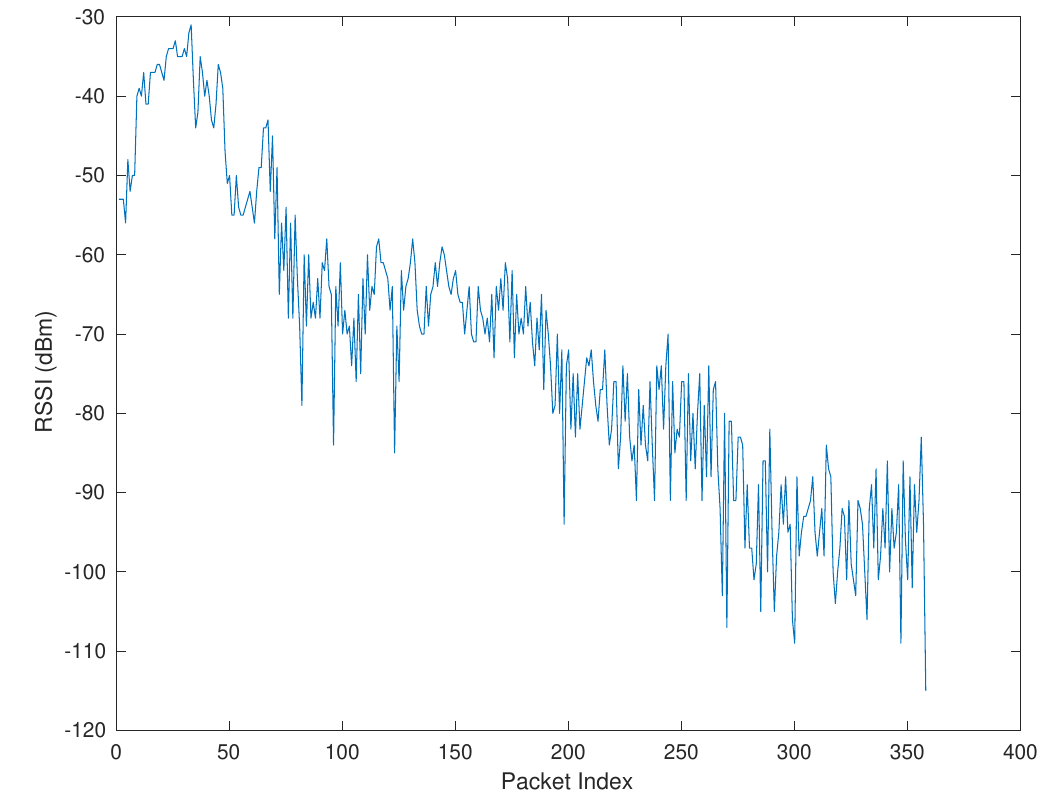}
	\caption{Link quality fluctuation between two communicating nodes as a result of the constant movement of the surface of water at one of our deployment settings. }
	\label{fig:crssi_cc1200}
\end{figure}

Figure \ref{fig:crssi_cc1200} displays the change in the RSSI of received packets. For this  deployment, the transmitter was deployed on the surface of water and the receiver, on dry land. A 25 mph North-East wind was blowing, thus moving the transmitter appreciably in a three-dimensional space. The RSSI consists of two components, a long-term variation, which is distance dependent (a result of the translational motion of the node), and a local variation, which is due to the local oscillation of the node. We can exploit these characteristics to estimate the RSSI at time $t$ in terms of its past history as well as the perceived change in the acceleration at time $t$: 
\begin{equation}
\label{eq:model_3}
    \hat{ \mathbf{r}}(t) =  \rho (t) \mathbf{r} (t-1) + \alpha (t) \mathbf{a} (t),
\end{equation}
where $\mathbf{a}(t)$ is the 3D acceleration the node experienced at time $t$. Both $\mathbf{r}(t)$ and $\mathbf{a}(t)$ are random variables. The objective of the minimum mean square error estimation is to minimize the  mean square error:% -- the difference between $\mathbf{r} (t)$ and $\hat{\mathbf{r}(t)}$:
\begin{equation}
\label{eq:model_4}
E \left [ \mathbf{e}^2(t) \right ] = E \left [ \left (\mathbf{r} (t) - \hat{\mathbf{r}}(t) \right )^2 \right ].
\end{equation}
The mean square error is minimized by determining the optimal $\rho(t)$ and $\alpha(t)$, which are, in turn, determined by inserting Equation~\ref{eq:model_3} in Equation~\ref{eq:model_4} and differentiating Equation~\ref{eq:model_4} with respect to $\rho(t)$ and $\alpha(t)$ and setting the results to 0:
\begin{align}
\label{eq:model_4_orthogonal}
\frac{\partial E \left [ \mathbf{e}^2(t)\right]}{\partial \rho(t)} & =   E \left [ \left ( \mathbf{r}(t) - \hat{\mathbf{r}}(t) \right ) \mathbf{r}(t-1) \right]  = 0 \\ \nonumber 
\frac{\partial E \left [ \mathbf{e}^2(t)\right]}{ \partial \alpha(t)} & =   E \left [ \left ( \mathbf{r}(t) - \hat{\mathbf{r}}(t) \right ) \mathbf{a}(t) \right]  = 0.  \nonumber 
\end{align}
Note that, due to the three-dimensional nature of the acceleration, $\alpha(t)$ is a compound term consisting of $\alpha_x(t)$,  $\alpha_y(t)$, and  $\alpha_z(t)$. Hence, we have\footnote{We shall drop the time indices whenever the context is clear.}:

\begin{footnotesize}
\begin{align}
\label{eq:model_5}
E \left [ \mathbf{r} (t) \mathbf{r} (t-1) \right ]  & = \rho E \left [ \mathbf{r}^2(t-1) \right ] + \dots + \alpha_z E \left [ \mathbf{r} (t -1) \mathbf{a}_z (t) \right] \\ \nonumber
E \left [ \mathbf{r} (t) \mathbf{a}_x (t) \right ]  & = \rho E \left [ \mathbf{r}(t-1) \mathbf{a}_x (t) \right ] + \dots + \alpha_z E \left [ \mathbf{a}_x (t) \mathbf{a}_z (t) \right] \\ \nonumber
E \left [ \mathbf{r} (t) \mathbf{a}_y (t) \right ]  & = \rho E \left [ \mathbf{r}(t-1) \mathbf{a}_y (t) \right ] + \dots + \alpha_z E \left [ \mathbf{a}^2_y (t) \mathbf{a}_z (t) \right] \\ \nonumber
E \left [ \mathbf{r} (t) \mathbf{a}_z (t) \right ]  & = \rho E \left [ \mathbf{r}(t-1) \mathbf{a}_z (t) \right ] + \dots + \alpha_z E \left [ \mathbf{a}^2_z (t) \right]. 
\end{align}
\end{footnotesize}
Or in matrix form we have:
\begin{equation}
    \label{eq:model_6b}
    \begin{bmatrix}
    R_{r} \\
    R_{x} \\
    R_{y} \\
    R_{z} 
    \end{bmatrix}
     = 
     \begin{bmatrix}
         R_{rr} & R_{rx} & R_{ry}  & R_{rz} \\ 
         R_{rx} & R_{xx} & R_{xy}  & R_{xz} \\ 
         R_{ry} & R_{xy} & R_{yy}  & R_{yz} \\ 
         R_{rz} & R_{xz} & R_{yz}  & R_{zz} \\ 
     \end{bmatrix}
     \begin{bmatrix}
         \rho \\
         \alpha_x \\
         \alpha_y \\
         \alpha_z \\
     \end{bmatrix},
\end{equation}
where $R_{r}$ refers to the correlation between $\mathbf{r}(t)$ and $\mathbf{r}(t-1)$; $R_{x}$ refers to the correlation between $\mathbf{r}(t)$ and $\mathbf{a}_x(t)$, etc.; $R_{rr}$ refers to the correlation of $\mathbf{r} (t-1) $ with itself, $R_{rx}$ refers to the correlation between $\mathbf{r}(t-1)$ and $\mathbf{a}_x$; and $R_{xy}$ refers to the correlation between $\mathbf{a}_x(t)$ and $\mathbf{a}_y(t)$, etc. The left term in Equation~\ref{eq:model_6b} encodes the correlation of the measurement with the parameter to be estimated, i.e., $\mathbf{r}(t)$; the middle term encodes the correlation of the measurement with itself; and the last term contains the coefficients which optimize the estimation. The strength of the correlations depends on many factors, including the synchronization between the samples representing the RSSI and the 3D acceleration; the magnitude and direction of the waves; the rates at which the RSSI and the inertial sensors are sampled; as well as the observation period. More compactly, we can express Equation~\ref{eq:model_6b} as:
\begin{equation}
    \label{eq:model_6}
     R =   E   A.
\end{equation}
%where $R$ refers to the correlation between the observation and the parameter to be estimated (i.e., the received power); $E$ refers to the correlation between the observation (evidence); and $A$ refers to the mean square coefficients to be estimated. Referring to the three terms in Equation~\ref{eq:model_6} as $R$, $E$, $A$, respectively, 
Hence, the optimal coefficients minimizing the mean square error in Equation~\ref{eq:model_4} can be expressed as follows:
\begin{equation}
\label{eq:model_7}
 A = E^{-1} R.
 \end{equation}

\subsection{Gradient Descent}
However, Equation~\ref{eq:model_7} relies on the inverse of $E$. To solve for $A$, we can use standard linear systems solvers from numerical analysis e.g., Cholesky decomposition \cite{trefethen2022numerical}. To solve the system $R=EA$, where $R$ is $m\times m$ positive definite matrix, costs $O(m^3)$ operations. An alternative light-weight approach is to employ gradient descent, which is the approach we adopt in this paper. Thus, instead of solving the system $R=EA$ directly, we consider the following optimization problem:

\begin{equation}
\label{eq:model_7a}
\min_{A}\,\,\, \frac{1}{2} \|E A - R\|_2^2,
\end{equation}
where $||\cdot||_2$ denotes the $\ell_2$ norm. Equivalently, we can solve
%Expanding the objective function, we have:
%\[
%\frac{1}{2} \|E A - R\|_2^2 = \frac{1}{2} A^T E^T E A - A^T E^T R + \frac{1}{2} R^T R.
%\]
%Since the last term does not depend on $A$, the problem is equivalent to minimizing:
%\[
%\min_{A}\,\,\, \frac{1}{2} A^T E^T E A - A^T E^T R.
%\]
\begin{equation}
\label{eq:model_7b}
\min_{A}\,\,\, \frac{1}{2} A^T E^T E A - A^T E^T R.
\end{equation}
To solve this minimization problem, we employ the gradient descent algorithm. Let $f(A) = \frac{1}{2} A^T E^T E A - A^T E^T R$. The negative gradient of this function is given by:
\begin{equation}
\label{eq:model_7c}
-\nabla f(A) = E^T R - E^T E A.
\end{equation}
Gradient descent starts with an initial estimate $A^{(0)}$ and iteratively moves in the direction of the negative gradient to minimize the function. The update rule is:
\begin{equation}
\label{eq:model_7d}
 A^{(k+1)} = A^{(k)} + \alpha_k r_k, \quad k = 0, 1, 2, \dots,
\end{equation}
where $\alpha_k$ denotes the step size at the $k$-th iteration, and $r_k = E^T R - E^T E A^{(k)}$. From gradient descent theory, the optimal step size $\alpha_k$ can be determined as:
\begin{equation}
\label{eq:model_7e}
\alpha_k = \frac{||r_k^T||_2^2}{r_k^T E^T E r_k}.
\end{equation}
We repeat the iterative updates until either a maximum number of iterations $T$ is reached or the norm of the gradient falls below a pre-defined threshold. Compared to solving the linear system directly, gradient descent is computationally efficient, relying only on matrix-vector products. Specifically, when $R$ is $m\times m$ positive definite  matrix, each iteration of gradient descent costs $O(m^2)$. A critical part of the algorithm is the initialization $A^{(0)}$. Running the algorithm for a few iterations provides reasonable solutions with ease of implementation, minimal computational and memory overhead.

\subsection{Model Statistics}

The inverse in Equation~\ref{eq:model_7} implies that $\rho$ is directly proportional to $E[\mathbf{r}(t) \mathbf{r}(t-1)]$ and inversely proportional to $E[\mathbf{r}^2(t-1)]$; similarly, $\alpha_x$ is directly proportional to $E[\mathbf{r}(t)\mathbf{a}_x(t)]$, but inversely proportional to $E[\mathbf{a}^2_x(t)]$, etc. This is logical, as our confidence in the estimation increases if the correlation between the parameter to be estimated and the measurement increases, but decreases if the variance of the measurement increases. Once the optimal coefficients are determined, it is possible to determine the mean and the variance of the error:
\begin{align}
\label{eq:model_8}
    E \left [\mathbf{e} (t) \right ]    & = E \left [ \mathbf{r}(t) - \hat{\mathbf{r}}(t)\right ] \\ \nonumber
                                    & =  E \left [ \mathbf{r}(t) - \left ( \rho(t) \mathbf{r}(t - 1) + \alpha (t)\mathbf{a}(t) \right ) \right ] \\ \nonumber
                                    & = E \left [ \mathbf{r}(t) \right] - \rho (t) E \left [ \mathbf{r}(t-1) \right ] - \alpha (t) E \left [\mathbf{a} (t) \right ]. 
\end{align}
In Equation~\ref{eq:model_8}, the acceleration is once again expressed in a compact form. In the absence of any acceleration, the difference between $\mathbf{r}(t)$ and $\mathbf{r}(t-1)$ is due to a measurement error only and, therefore, the estimation error is 0. From Equation~\ref{eq:model_4_orthogonal}, it is clear that the error is orthogonal to the measurement (data) (ref. also to \cite{Liu2023, guo2011estimation}, and \cite{papoulis2002probability}, pp. 261-269); hence, the optimal mean square error is:

\begin{small}
\begin{align} 
\label{eq:model_9}
 E \left [ \mathbf{e}^2 (t)\right ] & =  E \left [ \left ( \mathbf{r}(t) - \hat{\mathbf{r}}(t) \right ) \mathbf{r}(t) \right ] \\ \nonumber 
     & = E \left [ \mathbf{r}^2(t) \right ] - \rho (t) E \left [ \mathbf{r}(t) \mathbf{r} (t-1) \right]  + 
 \alpha (t) E \left [ \mathbf{r}(t)\mathbf{a}(t) \right ].
\end{align}
\end{small}
One of the most significant consequences of Equations~\ref{eq:model_8} and \ref{eq:model_9} is that the conditional probability density function of the change in the received power can be described in terms of them:

\begin{small}
\begin{align}
    \label{eq:model_10}
    f \left ( r(t)|r(t-1),a(t) \right )  =   \nonumber \\
    \frac{1}{\sqrt{2 \pi P}}  \exp \left \{- \frac{\left ( r(t) - \rho(t) r(t-1) - \alpha(t) a(t) \right)^2}{P} \right \},
\end{align}
\end{small}
where $P = E \left [ \mathbf{e}^2(t) \right ]$ as in Equation~\ref{eq:model_9}. Equation~\ref{eq:model_10} presupposes that both $\mathbf{r}(t-1)$ and $\mathbf{a}(t)$ are normally distributed, which they are, as we already demonstrated in Figs.~and \ref{fig:histogram_cc1200} and \ref{fig:histogram_cc2538}.
\section{Evaluation}
\label{sec:evaluation}

In our experiments, floating nodes measured the 3D motion they experienced using an onboard IMU and communicated this with a base station deployed on the ground. Up on receiving a packet, the base station associated the received power with the motion data. The packet size was 128 bytes. For this packet size, the sustainable rate the CC1200 radio could support was 2 packets per second, whereas the CC2538 radio could support 10 packets per second. These rates were highly environment dependent, including the external temperature and the internal heat dissipation. The waves at Miami South Beach were distinctly different from the waves at Crandon Beach. The former were typically bigger and longer, whereas the latter were typically shorter and faster and more harmonious. Since our estimation establishes a correlation between the RSSI and the linear acceleration, ideally, the two should be sampled at the same rate. However, sampling the IMU at a rate below 10 Hz resulted in a significant estimation error. Therefore, for all our experiments, the IMU was sampled at a rate of 10 Hz. For the estimations concerning the CC1200 radio, the measurements of the IMU were down-sampled using overlapping windows of size 10 samples and an overlapping coefficient of 0.5. Because the IMU and the RSSI were measured in different units, they were normalized using the max/min normalization, so that they had values between 0 and 1. 

\subsection{Prediction}
\begin{figure}
    \centering    \includegraphics[width=1\linewidth]{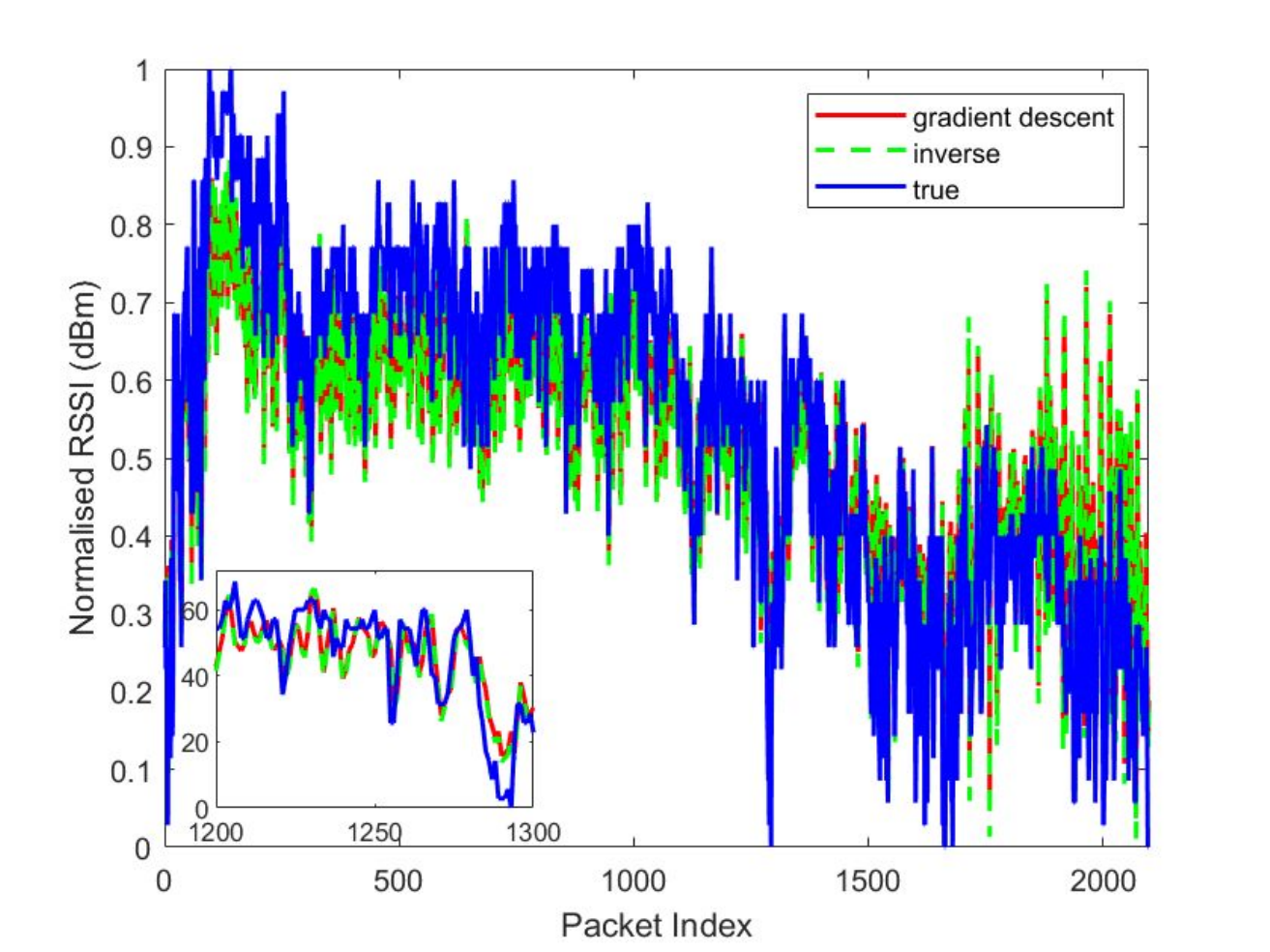}
    \caption{Difference between normalized actual RSSI, and estimated RSSI obtained by gradient descent and solving the exact linear system. The deployment is for Crandon Beach. Radio: CC2538.}
    \label{fig:rssi_comparison_cb}
\end{figure}

\begin{figure}
    \centering    \includegraphics[width=1\linewidth]{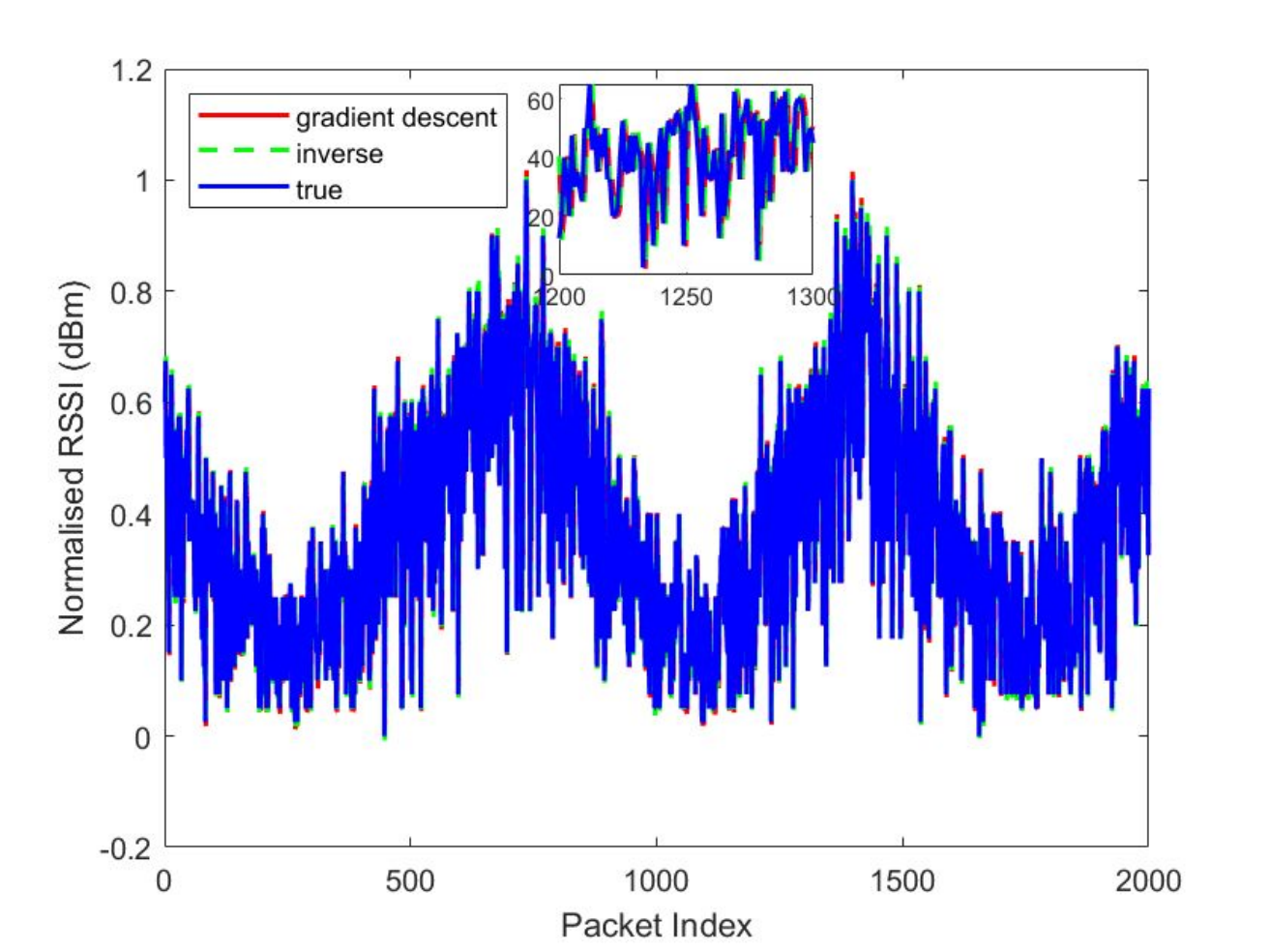}
    \caption{Difference between normalized actual RSSI, and estimated RSSI obtained by gradient descent and solving the exact linear system. The deployment is for Miami south Beach. Radio: CC2538.}
    \label{fig:rssi_comparison_sb}
\end{figure}

In this subsection, we compare the normalized actual (measured) RSSI with its estimates obtained using the model based on gradient descent and the one involving matrix inversion. For the gradient descent, we consider two scenarios: initial zero values to all entries of $A$ and random assignment from a uniform distribution over $([0, 1])$. Figures ~\ref{fig:rssi_comparison_cb} and \ref{fig:rssi_comparison_sb} present these comparisons for the deployments of Crandon Beach and Miami South Beach, respectively. The number of iterations for the gradient descent were 100 in both cases. The radio under consideration is the CC2538 System-on-Chip. As can be seen, for the Miami South Beach deployment, the estimates agree very well with the actual normalized RSSI. However, for the Crandon beach deployment, while the two estimates agree quite well, they do not fully capture the profile of the normalized actual RSSI.

\begin{figure}[tbh!]
	\centering
        \includegraphics[width=0.45\textwidth]{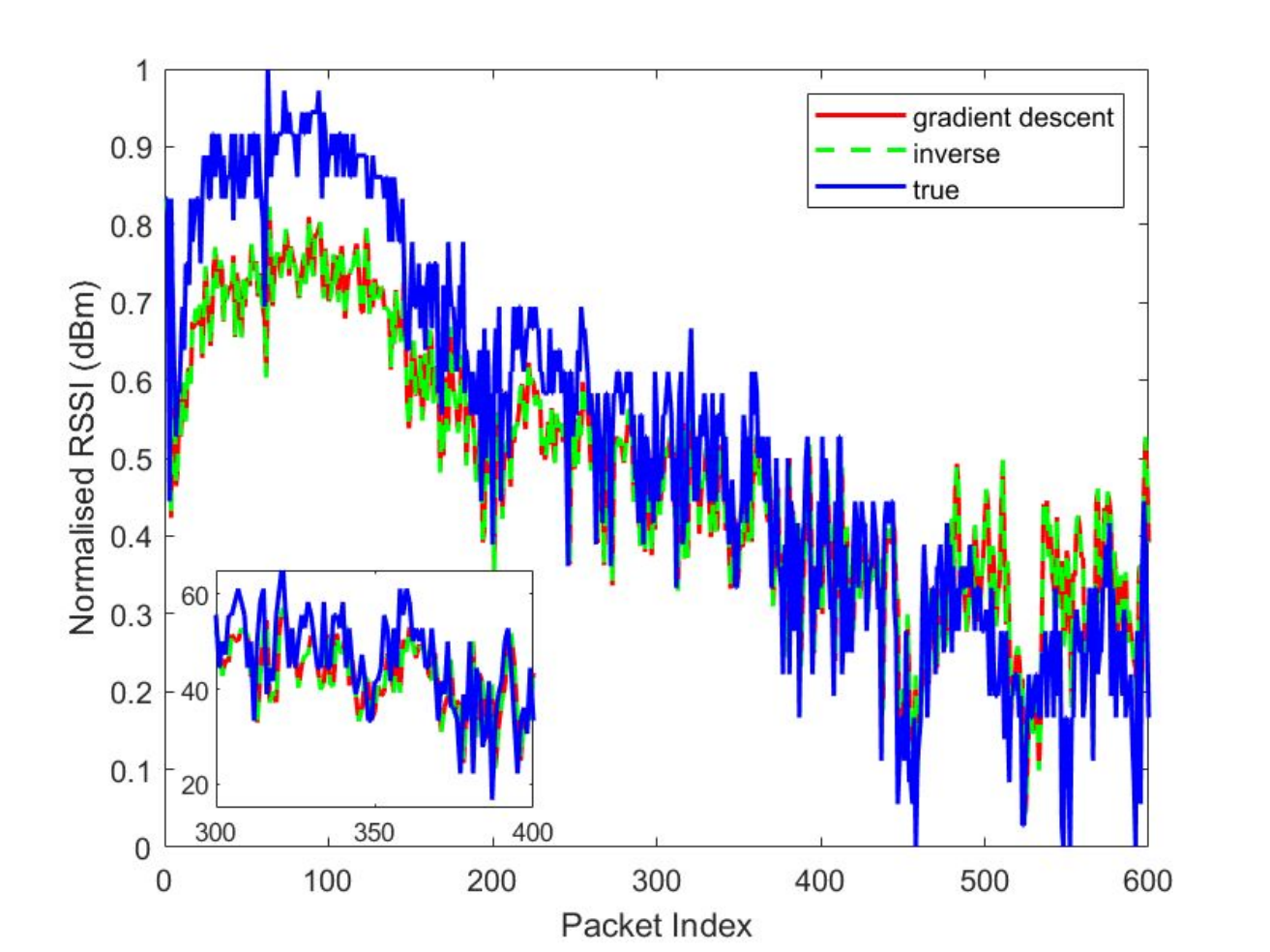}
	\caption{Difference between normalized actual RSSI, and estimated RSSI obtained by gradient descent and solving the exact linear system. The deployment is for Crandon Beach. Radio: CC1200.}
	\label{fig:estimation_mmse_cc1200_crando}
\end{figure}

Similarly, Figure \ref{fig:estimation_mmse_cc1200_crando} compares the actual RSSI with its estimates for the CC1200 radio. The deployment we consider is the Crandon Beach. We see that the two estimates based gradient descent and inversion agree very well, however they do not fully capture the profile of the normalized actual RSSI especially at low packet index.

\subsection{Root Mean Square Error}
Tables \ref{tab:mse_comparison_cb} and \ref{tab:mse_comparison_sb} compare the root mean square errors (RMSE) of the model based on gradient descent for different iterations. For the Crandon Beach deployment, with random initialization, we observe that even with just 10 iterations, the RMSE is comparable to that obtained by solving the exact linear system. For the Miami South Beach deployment, gradient descent performs better in terms of RMSE for both random and zero initialization, regardless of the total number of iterations. Tables \ref{tab:time_comparison_cb} and \ref{tab:time_comparison_sb} present the computation time of the different iterations for the Crandon Beach and Miami South Beach deployments, respectively. We note that gradient descent shows comparable runtime performance to solving the linear system. Overall, considering all the deployments and the experiments we carried out for both radios, the gradient-descent's prediction accuracy was on average 91\%, whereas the one involving matrix inversion was slightly worse, 90.7\%.

\begin{table}[h!]
    \centering
    \caption{Comparison of the RMSE of two different initialization using gradient descent. The mean square error using the exact linear system is $0.115$. Deployment: Miami Crandon Beach. Radio: CC2538.}
    \label{tab:mse_comparison_cb}
    \begin{tabular}{|c|c|c|}
        \hline
        $T$ & Mean Square Error  & Mean Square Error  \\ 
         (\# of Iterations) & (Zero initialization) & (Random initialization) \\ \hline
        1000 & 0.113 & 0.114 \\ \hline
        500  & 0.111 & 0.114 \\ \hline
        100  & 0.107 & 0.114 \\ \hline
        50   & 0.102 & 0.115 \\ \hline
        10   & 0.0964 & 0.121 \\ \hline
  \end{tabular}
\end{table}

   \begin{table}[h!]
    \centering
    \caption{Comparison of RMSE for two different initialization using gradient descent. The mean square error using the exact linear system is $0.155$. Deployment: Miami South Beach. Radio: CC2538.}
    \label{tab:mse_comparison_sb}
    \begin{tabular}{|c|c|c|}
        \hline
        $T$ & MMSE  & RMSE  \\ 
         (\# of Iterations) & (Zero initialization) & (Random initialization) \\ \hline     
        1000 & 0.155 & 0.155 \\ \hline
        500  & 0.155 & 0.155 \\ \hline
        100  & 0.155 & 0.154 \\ \hline
        50   & 0.154 & 0.153 \\ \hline
        10   & 0.153 & 0.151 \\ \hline
    \end{tabular}
\end{table}

\begin{table}[h!]
    \centering
    \caption{Comparison of computation time to estimate $E$ via solving the system in Equation \ref{eq:model_7} versus using gradient descent. Deployment:  Crandon Beach. Radio: CC2538.}
    \label{tab:time_comparison_cb}
    \begin{tabular}{|c|c|c|}
        \hline
        $T$ & Time taken  & Time taken  \\ 
         (\# of Iterations) & (Exact solve) & (Gradient descent) \\ \hline
        1000 & $1.17\times 10^{-4}$ & $1.10\times 10^{-3}$ \\ \hline
        500  & $9.69\times 10^{-5}$ & $6.15\times 10^{-4}$ \\ \hline
        100  & $1.07\times 10^{-4}$ & $3.76\times 10^{-4}$ \\ \hline
        50   &  $8.68\times 10^{-5}$ & $3.51\times 10^{-4}$ \\ \hline
        10   & $4.79\times 10^{-5}$ & $6.65\times 10^{-5}$ \\ \hline
  \end{tabular}
\end{table}

\begin{table}[h!]
    \centering
    \caption{Comparison of computation time to estimate $E$ via solving the system in Equation \ref{eq:model_7} versus using gradient descent. Deployment:  South Beach. Radio: CC2538.}
    \label{tab:time_comparison_sb}
    \begin{tabular}{|c|c|c|}
        \hline
        $T$ & Time taken  & Time taken  \\ 
       (\# of Iterations) & (Exact solve) & (Gradient descent) \\ \hline
        1000 & $1.24\times 10^{-4}$ & $9.91\times 10^{-4}$ \\ \hline
        500  & $1.07\times 10^{-4}$ & $5.93\times 10^{-4}$ \\ \hline
        100  & $1.16\times 10^{-4}$ & $4.62\times 10^{-4}$ \\ \hline
        50   &  $1.38\times 10^{-4}$ & $4.20\times 10^{-4}$ \\ \hline
        10   & $1.17\times 10^{-4}$ & $3.90\times 10^{-4}$ \\ \hline
  \end{tabular}
\end{table}

\section{Comparison}
\label{sec:comp}

One of the disadvantages of our model is that establishing the motion statistics incurs a communication and computation cost. We employed the Kalman Filter to predict the received power in terms of its past history, without relying on the motion statistics, in order to compare the performance of the two models. The Kalman estimation requires measurement and process error statistics, both of which can be established and updated with a modest cost \cite{elsisi2023robust}. The measurement error can be established, prior to deployment, using statistical fingerprinting. Similarly, the process error statistics, which depend on the deployment environment and gradually changing, can be established after transmitting packets for a few minutes. Once these statistics are available, the Kalman filter mixes prediction and measurement values recursively.

\begin{figure}
	\centering
		\includegraphics[width=0.45\textwidth]{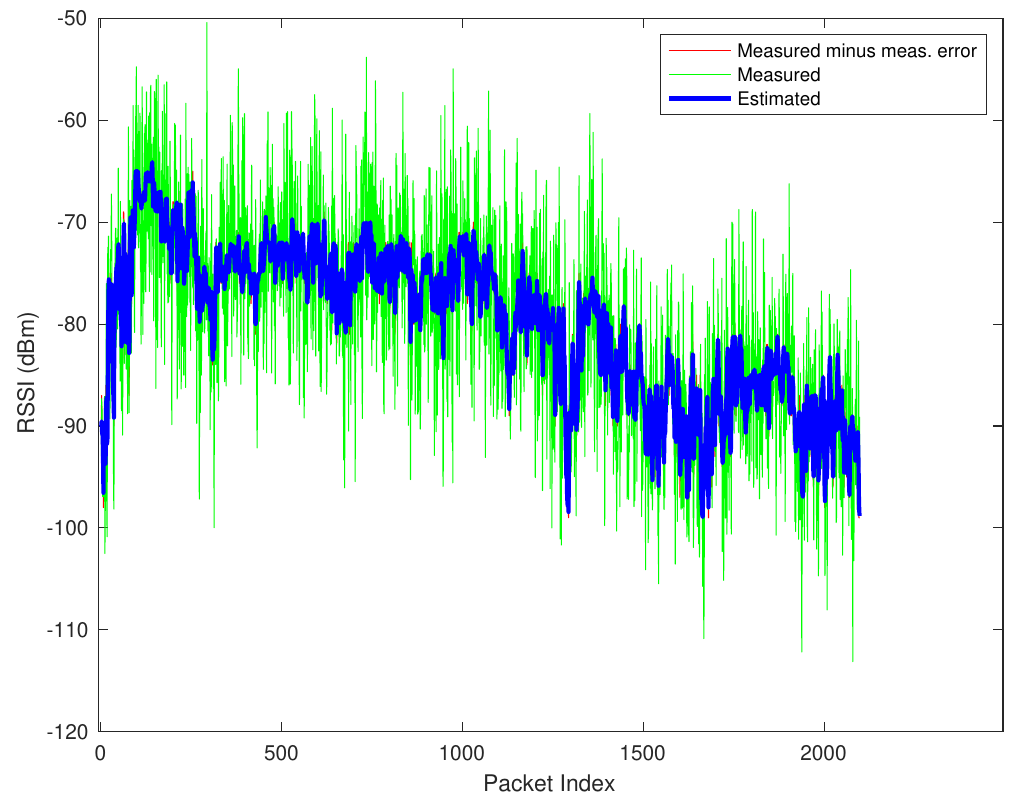}
	\caption{Link quality fluctuation estimated using Kalman Filter. Deployment: Miami Crandon Beach. Radio: CC2538.}
	\label{fig:estimation_kalman_cc2538_crandon}
\end{figure}

\begin{figure}
	\centering
		\includegraphics[width=0.450\textwidth]{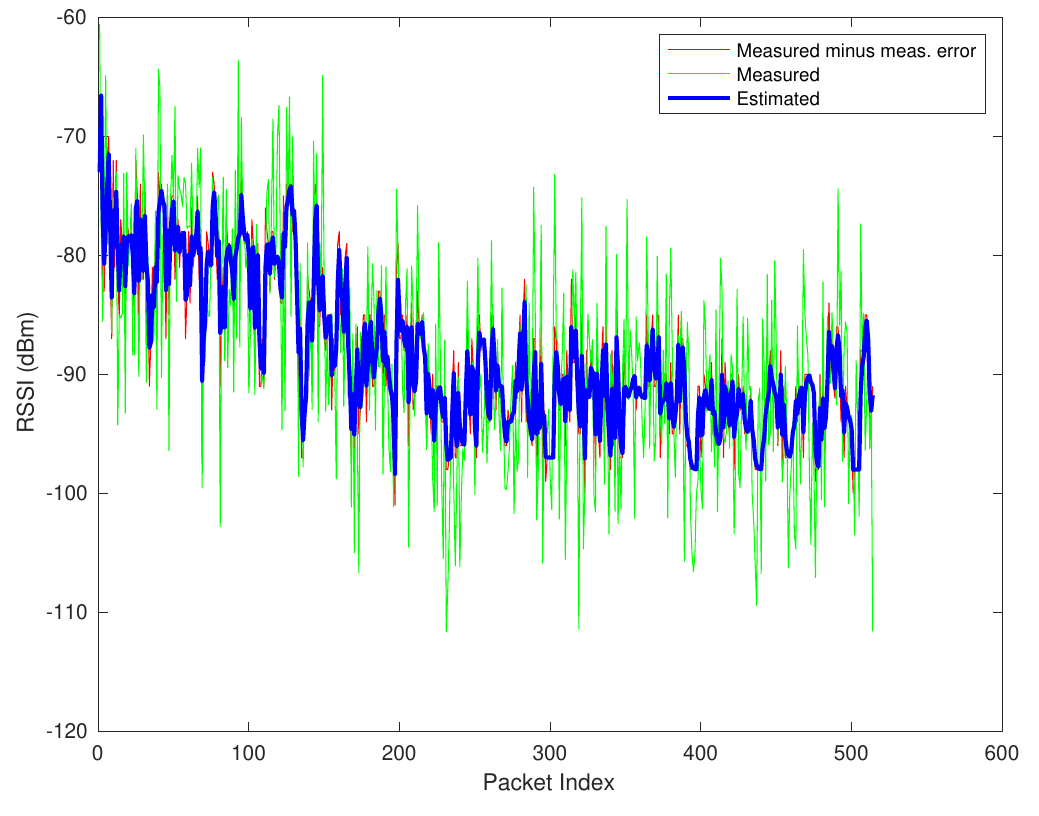}
	\caption{Link quality fluctuation estimated using Kalman Filter. Deployment: Miami South Beach. Radio: CC2538.}
	\label{fig:estimation_kalman_cc2538_sb}
\end{figure}

\begin{figure}
	\centering
	\includegraphics[width=0.23\textwidth]{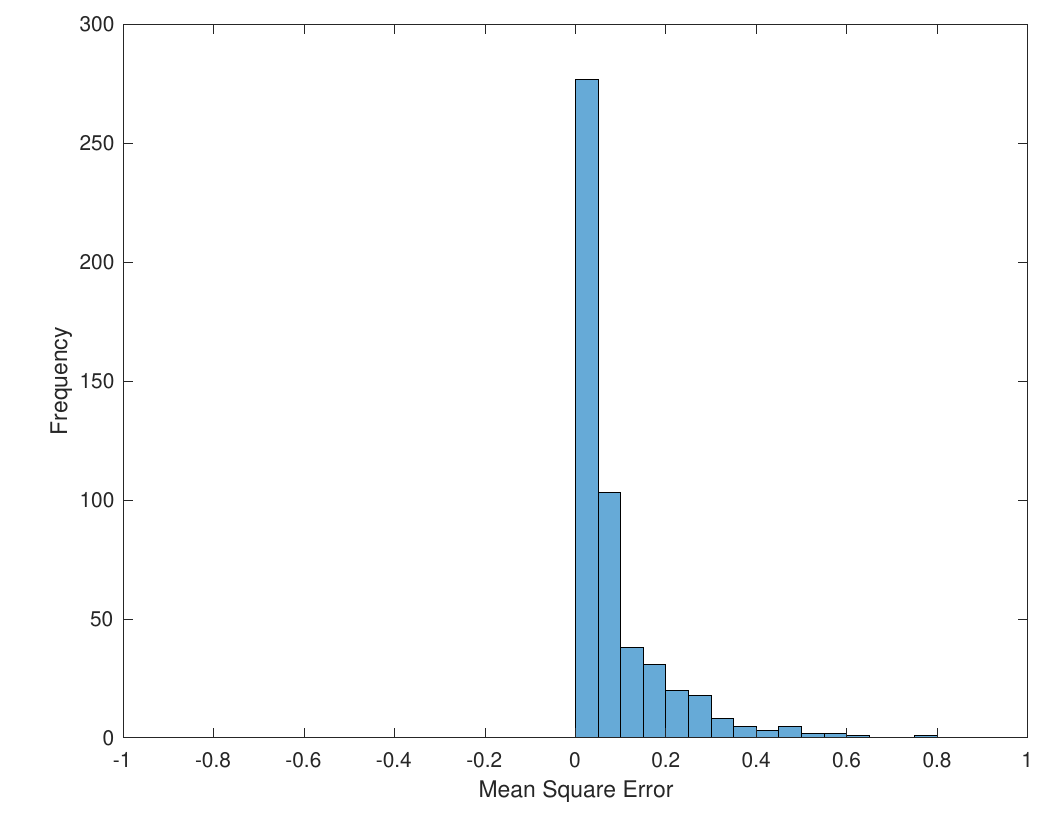}
	\includegraphics[width=0.23\textwidth]{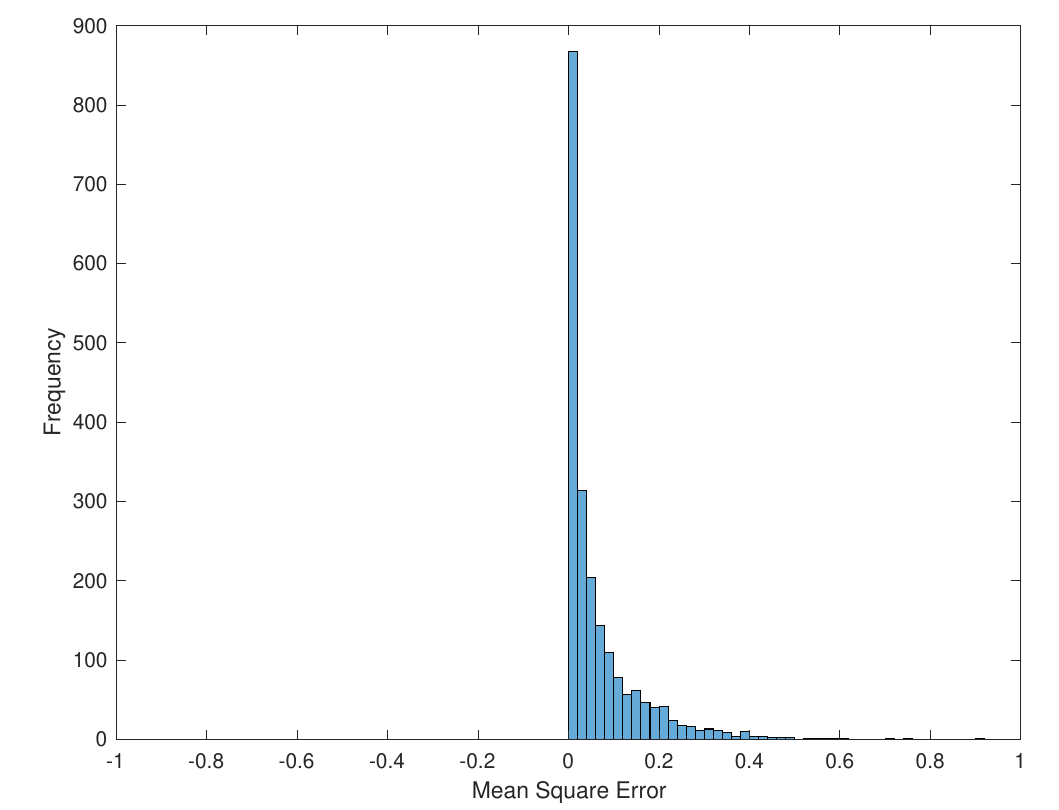}\caption{The histogram of the Minimum Mean Square Error for the Kalman Estimation. Left: South Beach. Right: Crandon Beach. Radio: CC2538}
	\label{fig:error_kalman_cc2538}
\end{figure}

Figures ~\ref{fig:estimation_kalman_cc2538_crandon} and~\ref{fig:estimation_kalman_cc2538_sb} display the Kalman estimation for two of the deployments we carried out at Crandon Beach and Miami South Beach using the CC2538 radio. The corresponding error statistics are displayed in Figure ~\ref{fig:error_kalman_cc2538}. For both cases, the average error is approximately 10\% and the variances of the mean square errors are slightly bigger. By comparison, the use of acceleration statistics in our estimation model improved the estimation accuracy by about 1\%. 

%\begin{figure}
%	\centering
%		\includegraphics[width=0.450\textwidth]{img/error_kalman_cc2538_cb.pdf}
%	\caption{The Histogram of the Minimum Mean Square Error for the estimation displayed in Fig.~\ref{fig:estimation_kalman_cc2538_crandon}.}
%	\label{fig:error_kalman_cc2538_cb}
%\end{figure}

\section{Conclusion} % 1
\label{sec:conclusion}

In this paper we proposed a model to predict the received power of low-power wireless sensing nodes deployed on the surface of rough waters. This is useful for adapting the transmission power of outgoing packets. We expressed the received power in terms of its past statistics and the statistics of the 3D motion the nodes experience using MMSE. However, MMSE involves matrix inversion, which is difficult to compute with resource-constrained devices. We avoided this stage by estimating the model parameters using the gradient-descent approach, which produced acceptable results even for a few number of iterations. Based on several experiments we carried out at South Beach Miami and Crandon Beach Miami using two different low-power radios, our approach predicted the received power with an average accuracy of ca. 91\%. By comparison, a Kalman Filter which relied only on the statistics of received power accomplished the same job with an average accuracy of 90\%. One of the challenges we faced during the modeling process was perfectly synchronizing the received power with the motion data. This might have resulted in performance penalty, since it has a direct bearing on Equation~\ref{eq:model_5}. In future, we aim to address this issue.

\balance
\bibliographystyle{IEEEtran}
\bibliography{library}

\end{document}